\documentclass[11pt]{article}

\usepackage[a4paper,margin=1in]{geometry}
\usepackage{amsmath,amssymb,amsfonts}
\usepackage{amsthm}
\usepackage{mathrsfs}
\usepackage{bm}
\usepackage{hyperref}
\usepackage{enumerate}
\usepackage{cite}

\usepackage{graphicx}
\usepackage{caption}
\usepackage{subcaption}

\theoremstyle{plain}

\theoremstyle{definition}

\theoremstyle{remark}

\begin{document}
	
	\title{\textbf{Option Pricing beyond Black-Scholes Model:} \\
		Quantum Mechanics Approach}
	\author{
	Pengpeng Li   \   and\   %\\
	Shi-Dong Liang\thanks{stslsd@mail.sysu.edu.cn}\\	
		\small School of Physics, Sun Yat-Sen University\\
		\small Guangzhou, China
	}
	
	\date{\today}

	\maketitle
	
	% =========================
	% 摘要
	% =========================
	\begin{abstract}
Based on the analog between the stochastic dynamics and quantum harmonic oscillator, we
propose a market force driving model to generalize the Black-Scholes model in finance market. We give new schemes of option pricing, in which we can take various unexpected
market behaviors into account to modify the option pricing. As examples, we present several market forces to analyze their effects on the option pricing. These results provide us two practical applications. One is to be used as a new scheme of option pricing when we can predict some hidden market forces or behaviors emerging. The other implies
the existence of some risk premium when some unexpected forces emerge.
	\end{abstract}
	
	\tableofcontents
	
	% =========================
	\section{Introduction}
	% =========================
	
The option pricing is a crucial issue in finance.
The Black-Scholes model gives a guideline to price options by the risk-free scheme,
which assumes the portfolio satisfies the no-arbitrage condition, perfectly hedge,
invariant interests, no transaction cost and the continuous evolution of prices. \cite{01} However, recently one discovers that above assumptions cannot hold in the practical finance markets, such as inconstant the risk-free interest rate, or non-continuously evolution, and fluctuate  volatility,\cite{02} which implies that the log return distribution (return is equal to the future price minus the original price) deviates normal (Guassian) distribution.
These phenomena induce a lot of interests to modify the Black-Scholes method.\cite{03,04,05,06,07} Belal E. Baaquie proposed a path integral method to optimize the evaluation of path-dependent options.\cite{03} By an elasticity variance model, Beni Lauterbach and Paul Schultz take the variant interest into account to give a new scheme of price.\cite{04}
Moreover, Louis O. Scott proves that the accurate option prices can be computed via Monte Carlo simulations when the variance changes randomly.\cite{04,05} Interestingly,
H. Kleinert and J. Korbel claim that the prices of options can also be evaluated by the double-fractional differential equation and its solution provide a more reliable hedge comparing with Black-Scholes formula.\cite{06}
Further, Lina Song and Weiguo Wang optimizes the fractional Black -Scholes Option pricing model by Finite Difference Method to give the solution of the difference equation.\cite{07} More recently, Robert C. Merton generalized the stock return distribution to
give an option pricing formula for the discontinuous returns.\cite{08}
Most importantly, considering the price distribution, R. N. Mantegna uses Levy Walk instead of original random walk and he gets a new price distribution deviating from normal distribution which can be applied into Black -Scholes Model. \cite{09}
These studies on the generalized Black-Scholes model provide
many ways to improve the Black-Scholes model and to make the option price close the realistic finance market.

In fact, there exist some hidden market forces, such as shorting or buying an underlying asset in finance market, which drive the dynamics of finance market and make the stochastic process of the finance market deviate Guassian distribution.
This non-Guassian effect should modify the option pricing.
Therefore, in this paper, we will propose the market-force concepts to
describe the stochastic dynamics of finance market based on the quantum mechanics approach.
The stochastic dynamics of finance market is described by the wave function, which
follows the Schr\"{o}dinger equation. The hidden market forces as
the market potential drive the stochastic dynamics of finance market, which
make the dynamics deviate the Guassian process and modify the Black-Scholes model which gives several schemes of option pricing.
In Sec. II, we present the market-force model of option pricing based on quantum mechanics.
In Sec. III, we propose several schemes of option pricing based on this model and discuss
their advantages and financial meanings. Finally we give the summary and conclusions
in Sec. IV.

\section{Market force Model of Stochastic dynamics}
\subsection{Black-Scholes model}
The dynamics of finance market is a stochastic process. The efficient market theory
assumes that there does not exist the arbitrage space, which implies that the
stochastic dynamics process is a Guassian process. The option pricing
of Black-Schole theory is based on the efficient market theory and the scheme
of option pricing is assumed to be risk free. The European call and put options are priced by
\begin{eqnarray}\label{BSf1}
	c&=& S_0 N(d_+)-Ke^{-rT}N(d_-)\\
	p&=& Ke^{-rT}N(-d_-)-S_0 N(-d_+)
\end{eqnarray}
where
\begin{equation}\label{BSd1}
	d_\pm = \frac{ln(S_0/K)+(r\pm\sigma^2/2)T}{\sigma\sqrt{T}}
\end{equation}
and $d_{-}=d_{+}-\sigma\sqrt{T}$.
$S_0$ is the current price of stock or asset and their corresponding delivery (strike) price $K$ if the option is exercised. $r$ is the risk-free rate and $\sigma$ is the volatility of asset. $T$ is the time to maturity of the option.
$N(d_{\pm})$ is the cumulative distribution function, which is expressed as
\begin{equation}\label{BSN1}
	N(d_{\pm})=\frac{1}{\sqrt{2\pi}}\int_{-\infty}^{d_{\pm}} e^{-\frac{x^2}{2}} dx
\end{equation}
where $N(d_{-})$ is the probability for the call option exercised in a risk-free world.
The expression $S_0 N(d_+)e^{-rT}$ is the expected stock price at
time $T$ in a risk-free world when stock prices less than the strike price are counted as
zero.

It can be seen that the cumulative distribution function plays a role of
probability in the risk-free market. When we consider a finance market driven by
some market force, shorting or buying, the Guassian probability distribution
of $N(d_{\pm})$ can be generalized to non-Guassian probability
distribution. We look for some hints from quantum mechanics how option pricing
work in a non-Guassian dynamics.

\subsection{Analog between finance market and quantum harmonic oscillator}
In general, the evolution of finance market is a stochastic dynamical process.
The Black-Scholes theory provides an option pricing scheme in a risk-free world,
which implies that the dynamical process of finance market is a Guassian process.
In quantum world the quantum state emerges also by a stochastic dynamics, in which
the probability density is expressed in terms of the norm of wave function.
The wave function evolution is driven by the Schr\"odinger equation. Therefore,
we can find an analog between the finance market and quantum mechanics.

(1) The finance market corresponds to the quantum bound systems.

(2) The stochastic dynamics of finance market corresponds to the dynamics of the quantum bound systems.

(3) The hidden shorting or buying in finance market corresponds to the intrinsic force or potential in quantum bound systems.

(4) The integrand function $\frac{1}{\sqrt{2\pi}}e^{-x^{2}/2}\equiv P_{BS}(x)$ of $N(d_{\pm})$ corresponds to the probability density $P(x)=|\psi(x)|^{2}$ of the quantum bound systems. Further, the Black-Scholes model corresponds to the ground state of quantum harmonic oscillator, namely, $P_{BS}(x)=P_{g,HO}(x)$. (See the following demonstration)

(5) The cumulative distribution function of the Black-Scholes model can be written as
\begin{equation}\label{BSN2}
	N(d_{\pm})=\int_{-\infty}^{d_{\pm}} P_{BS}(x) dx
	=\int_{-\infty}^{d_{\pm}} P_{g,HO}(x) dx
\end{equation}

This analog between finance market and quantum mechanics provides a
way to modify the integrand function $P_{BS}(x)$ of the Black-Scholes model
and give some new schemes of option pricing. Notice that this integrand function has the same form of the ground state wave function of quantum harmonic oscillator,
we start from the one-dimensional quantum harmonic oscillator.
The potential of the harmonic oscillator is
\begin{equation}\label{QHOP1}
	V(x)=\frac{1}{2}m\omega^{2}x^{2}
\end{equation}
and the stationary Schr\"{o}dinger equation is written as
\begin{equation}\label{QHO1}
	\frac{\hbar^2}{2m}\frac{d^2\psi}{dx^2}+\left(E-\frac{m\omega^2}{2}x^2\right)\psi=0
\end{equation}
The wave function in the ground state can be solved
\begin{equation}\label{QHOS1}
	\psi_{g}(x)=\left(\frac{\alpha}{\sqrt{\pi}}\right)^{1/2} e^{-\alpha^{2}x^{2}/2}
\end{equation}
where $\alpha=\sqrt{\frac{m\omega}{\hbar}}$.
For convenience we set $\alpha=\frac{1}{\sqrt{2}}$ such that the probability density in the ground state
\begin{equation}\label{PP1}
	P_{g,HO}(x)=\frac{1}{\sqrt{2\pi}} e^{-x^{2}/2}=P_{BS}(x)
\end{equation}
Therefore, the ground state of quantum harmonic oscillator corresponds to Black-Schole model.
This correspondence between finance market and quantum harmonic oscillator provides a way
to generalize the Black-Scholes model based on quantum mechanics approach.
When we add some forces to generalize the quantum harmonic oscillator,
the ground state of wave function and its corresponding probability density deviates the Guassian form. This infers that some market forces emergence makes the finance market deviate the Guassian process and modify the Black-Scholes option pricing.

The finance market force is defined by
\begin{equation}
	F=-\frac{dV(x)}{dx}
	\label{5}
\end{equation}
where $V(x)$ is the potential of the bound systems. General speaking,
force describes any local or individual behavior making finance market
deviate the equilibrium state, while potential describes the global effect
induced from these local or individual behaviors.
Thus, the finance market force describe the behaviors of shorting or buying the underlying asset or any economic news and psychological behaviors in finance market. $F=-\frac{dV(x)}{dx}>0$ means any market force pushing the asset or stock price high. $F=-\frac{dV(x)}{dx}<0$ means any market resistance bringing down the asset or stock price.

It should be pointed out that the finance market forces we introduce from quantum mechanics
will modify the option pricing from two ways. One is to modify $P_{g,HO}$, which means
to modify $P_{BS}$ and the cumulative distribution function, the other is to modify
the volatility $\sigma$ because the force drives the probability distribution deviating
from the Guassian distribution. The effective volatility can be obtained by
\begin{equation}\label{Sigma}
	\sigma_{\textrm{eff}}=\sigma \sigma_{QM}
\end{equation}
where $\sigma_{QM}=\sqrt{\left\langle x^{2}\right\rangle_{QM}-\langle x\rangle_{QM}^{2}}$ and $\left\langle f(x) \right\rangle_{QM}=\int f(x) P_{QM}dx$, where $P_{QM}$ is the probability density from quantum mechanics. The volatility $\sigma_{QM}$ in the Black-Scholes formula $(\ref{BSf1})$ is $1$ for the standard Guassian distribution.

In principle, we can design different forces to study or describe different market behaviors and modify the option pricing.
When the force vanishes $F=0$, the quantum system reduces to the harmonic oscillator and our
model reduces to the Black-Scholes Model.
Thus, the standard harmonic oscillator potential $\frac{1}{2}m\omega^2 x^2$ can be regarded as the natural boundary condition of finance market.

Based on this analog between the finance market and quantum harmonic oscillator, we can
take different forces into account for the harmonic oscillator to generalize
the Black-Scholes option pricing for understanding their financial meaning.

\section{Market forces and option pricing}

\subsection{Constant forces}
\begin{figure}[h]
	\centering
	\includegraphics[scale=0.35]{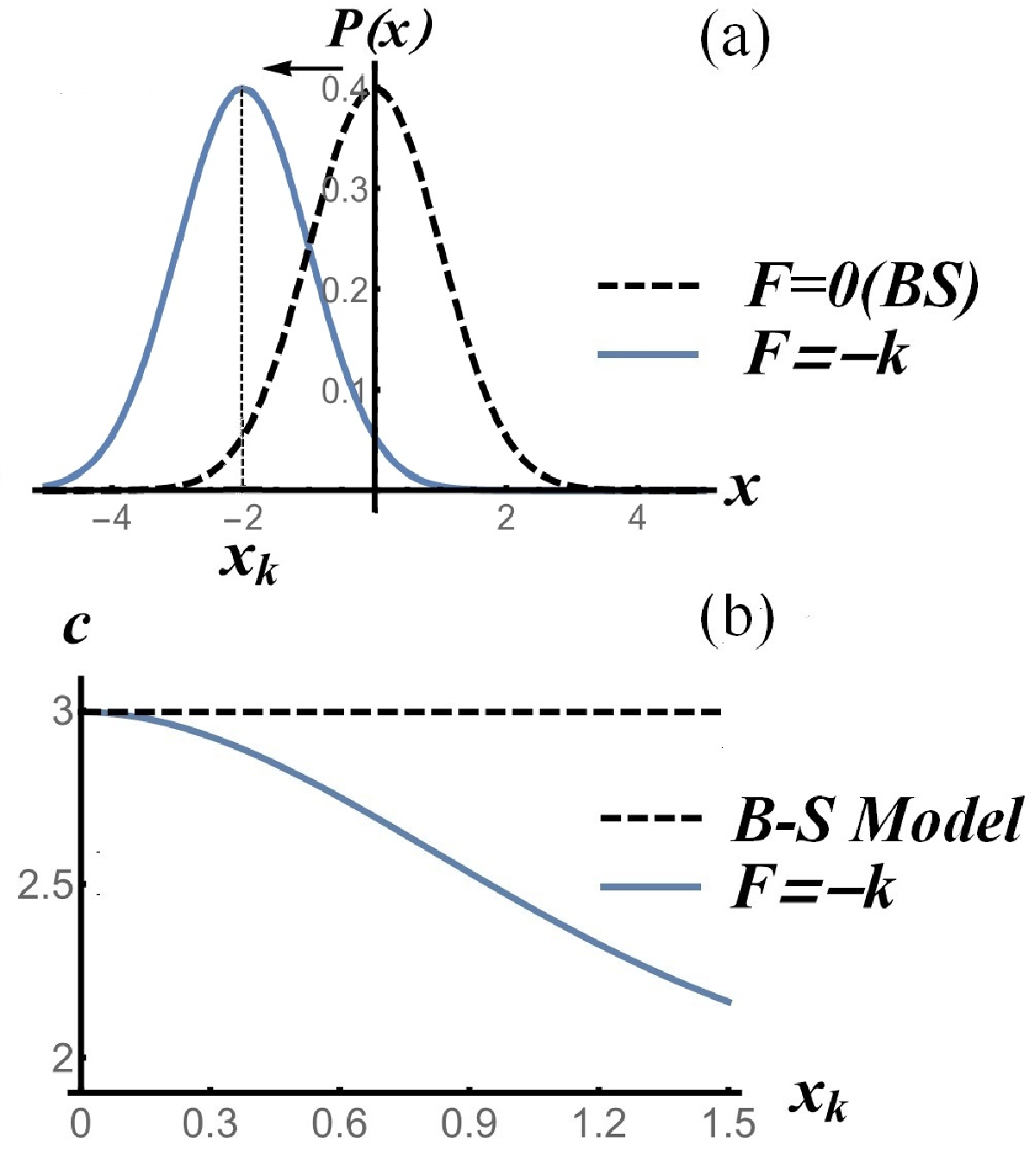}
	\caption{(Color online)(a) Comparison of the probability densities $P(x)$ between the normal and the force-driven probability densities. The solid line
		of the probability density driven by a constant force and its shift depends on the
		strength of force $k$.
		(b) The dashed line is the option price for $F=0$ (Black-Scholes model)
		The solid line is the option price modified by a constant force case $F<0$.
		The variance of call option price along with $x_{k}$.
		The parameters we use are $r=10\%$, $S_0=20$, $K=20$, $T=1$year and $\sigma=25\%$.
	}
	\label{fig1}
\end{figure}

Let us consider the market force be a constant $F=-k$. It corresponds to the potential
\begin{equation}\label{P1}
	V_{k}(x)=kx
\end{equation}
where $k$ is a small parameter describing the strength of the potential. Thus, the
Schr\"{o}dinger equation of system can be written as
\begin{equation}\label{QHO1}
	\frac{\hbar^2}{2m}\frac{d^2\psi}{dx^2}+\left(E-\frac{m\omega^2}{2}x^2
	+kx\right)\psi=0
\end{equation}
By solving the Schr\"{o}dinger equation, we obtain the solution of the wave function
in the ground state
\begin{equation}\label{QHOS2}
	\psi_{g}(x)=\left(\frac{1}{\sqrt{2\pi}}\right)^{1/2} e^{-(x-x_{k})^{2}/4}
\end{equation}
where $x_{k}=\frac{2k}{\hbar\omega}$.
The peak of the probability density is shifted to $x_{k}$ shown in Fig.$\ref{fig1}(a)$. Namely
\begin{equation}\label{PPk}
	P_{k}(x)=\frac{1}{2\pi}e^{-(x+x_{k})^2/2}
\end{equation}
Using the corresponding relation in
Eq. ($\ref{BSN2}$) we plot numerically the call option price versus the shift $x_{k}$ in Fig. $\ref{fig1}(b)$.

It can be seen from Fig. $\ref{fig1}$(a) that the shape of the probability density does not change and it moves to the left for the constant market force $F<0$. Similarly, it will move to the right for $F>0$. From the financial point of views, $F<0$
means that there exists shorting the asset in finance market and $F>0$ means buying behaviors in finance market. The option prices are shown in Fig. $\ref{fig1}$(b), in which the
the dashed line is the option price for $F=0$ (Black-Scholes model) and
the solid line for the constant force case $F<0$. When the force increases with $k$
the call option price should decrease monotonically with $k$ for $F<0$.
It matches the market behavior that sorting the underlying asset will reduce its call option price. Similarly it should increase with $k$ for $F>0$.

\subsection{Linear forces}
When we consider the market force being proportional to $x$, $F=-2\lambda x$, which induces the potential in finance market
\begin{equation}\label{P2}
	V_{\lambda}(x)=\lambda x^2
\end{equation}
where $\lambda>0$. The Schr\"{o}dinger equation of system can be written as
\begin{equation}\label{QHO1}
	\frac{\hbar^2}{2m}\frac{d^2\psi}{dx^2}+\left[E-\left(\frac{m\omega^2}{2}+\lambda\right)x^2
	\right]\psi=0
\end{equation}
The wave function in the ground state can be solved by
$\psi_{g}(x)=\left(\frac{\lambda_{\omega}}{\sqrt{2\pi}}\right)^{1/2}
e^{-\lambda_{\omega}x^{2}/4}$
and the probability density is
\begin{equation}\label{P3}
	P_{\lambda}(x)=\frac{\lambda_{\omega}}{\sqrt{2\pi}}e^{-\lambda_{\omega}x^{2}/2}
\end{equation}
where $\lambda_{\omega}=\sqrt{1+\frac{\lambda}{\omega}}$.

\begin{figure}[h]
	\centering
	\includegraphics[scale=0.4]{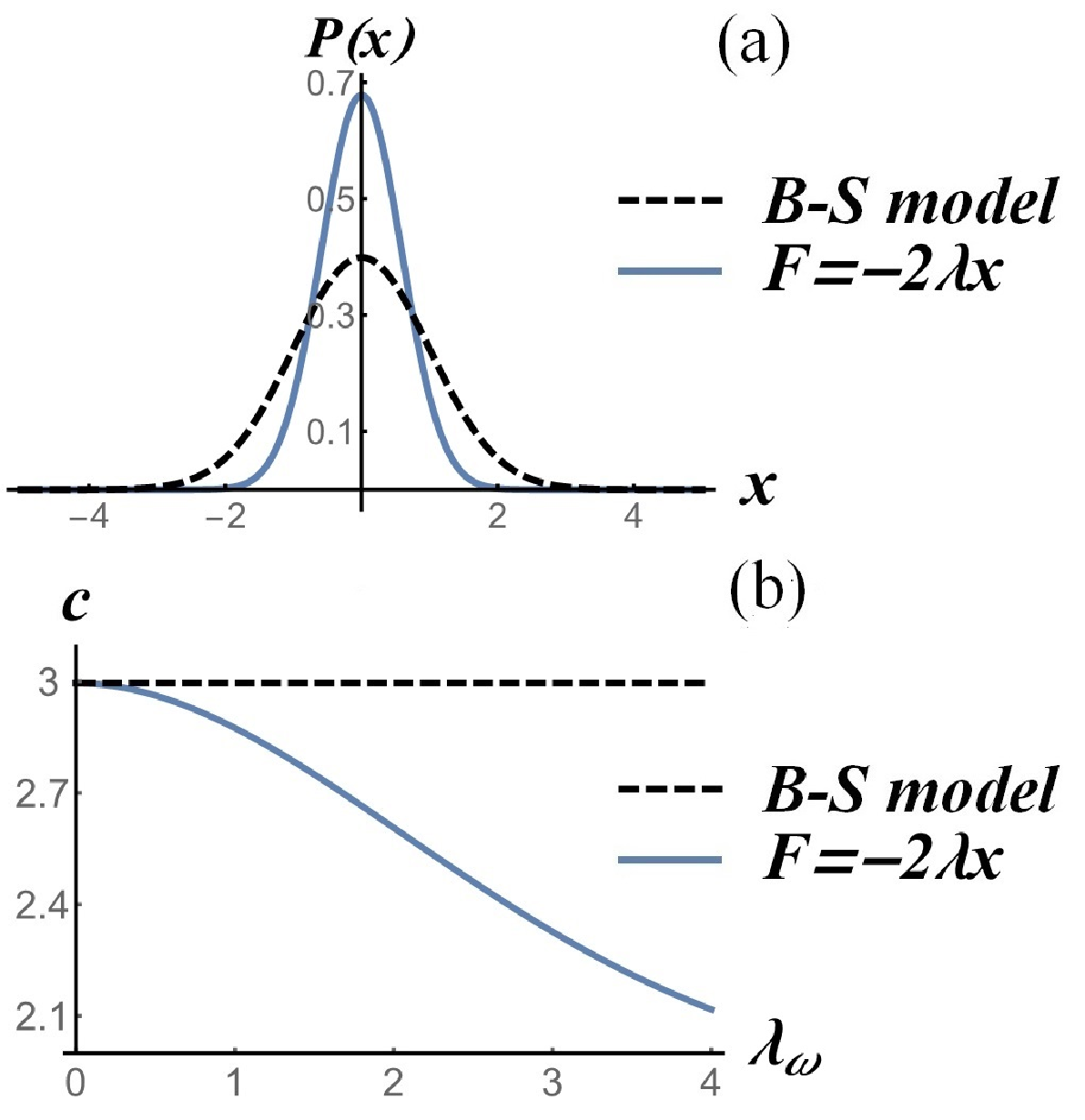}
	\caption{(Color online)(a)
		Comparison of the probability densities $P(x)$ between the normal and the force-driven probability densities. The solid line of the probability density driven by a $-2\lambda x$
		force. The volatility vary with the strength of force $\lambda$.
		(b) The dashed line is the option price for $F=0$ (Black-Scholes model)
		The solid line is the option price modified by the force.
		The parameters we use are $r=10\%$, $S_0=20$, $K=20$, $T=1year$ and $\sigma=25\%$.
	}
	\label{fig2}
\end{figure}

Figure $\ref{fig2}$(a) shows the probability densities of the Black-Scholes model
and the generalized harmonic oscillator with the linear force.
It can be seen that the force modifies the peak height and the volatility of the probability density such that the option prices vary with the force strength seen in Fig. $\ref{fig2}$(b).

\subsection{$x^{2}$ forces}
For the market force $F=-3\beta x^2$, which induces the
potential in finance market
\begin{equation}\label{P3}
	V_{\beta}(x)=\beta x^3
\end{equation}
The Hamiltonian for this case can be written as
\begin{equation}\label{HP3}
	\widehat{H}=\widehat{H}_{0}+V_{\beta}(x)
\end{equation}
For small $\beta$, we can use the perturbation method to turn out the approximate
solution of wave function in the ground state. The first-order approximation solution is zero. The second-order approximation solution of wave function
can be obtained by (see Appendix)
\begin{equation}\label{WFP3}
	\psi_{g}(x)=\frac{C}{(2\pi)^{1/4}}e^{-x^2/4}
	\left[1-\frac{\beta}{\hbar\omega}\left(\frac{1}{3}x^3+\frac{x}{2}\right)\right]
\end{equation}
The probability density is obtained
\begin{equation}\label{PDP3}
	P_{\beta}(x)=C^2\frac{e^{-x^2/2}}{\sqrt{2\pi}}
	\left[1-\frac{\beta}{\hbar\omega}\left(\frac{1}{3}x^3+\frac{x}{2}\right)\right]^{2}
\end{equation}
where $C$ is the normalization constant.

\begin{figure}[h]
	\centering
	\includegraphics[scale=0.4]{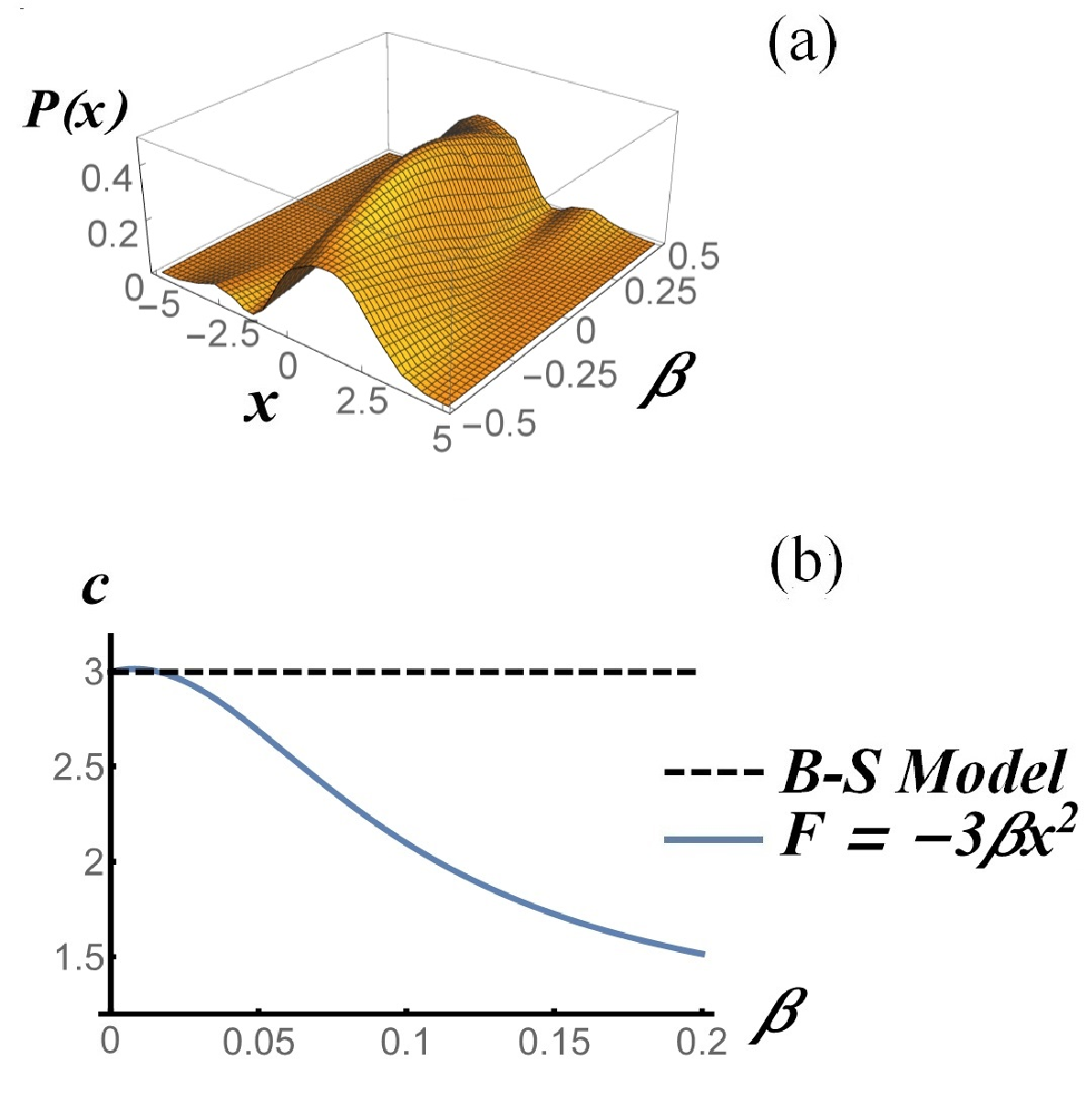}
	\caption{(Color online)(a) The probability density versus $\beta$
		When $\beta=0$ it reduces the normal distribution assumed in Black-Scholes Model.
		(b) The call option price versus $\beta$.
		The parameters we use are $r=10\%$, $S_0=20$, $K=20$, $T=1year$ and $\sigma=25\%$.
	}
	\label{fig3}
\end{figure}

We plot the probability density versus $\beta$ in Fig. $\ref{fig3}$(a) and the option price in Fig. $\ref{fig3}$(b).

We can see from Fig. $\ref{fig3}$(a) that the main peak moves to right. As $\beta$ increase a sub-peak appears. This phenomena can be interpreted by the collective effect in finance market in the natural boundary condition.
Fig. $\ref{fig3}$(b) shows the option price with $\beta$. As $\beta$ increases the option price goes down with shorting force, which matches our expectation.

\subsection{$x^{3}$ forces}
Further we can consider the market force $F=-4\gamma x^3$, which induces the
potential in finance market
\begin{equation}\label{P4}
	V_{\gamma}(x)=\gamma x^4
\end{equation}
The Hamiltonian for this case can be written as
\begin{equation}\label{HP3}
	\widehat{H}=\widehat{H}_{0}+V_{\gamma}(x)
\end{equation}

\begin{figure}[h]
	\centering
	\includegraphics[scale=0.4]{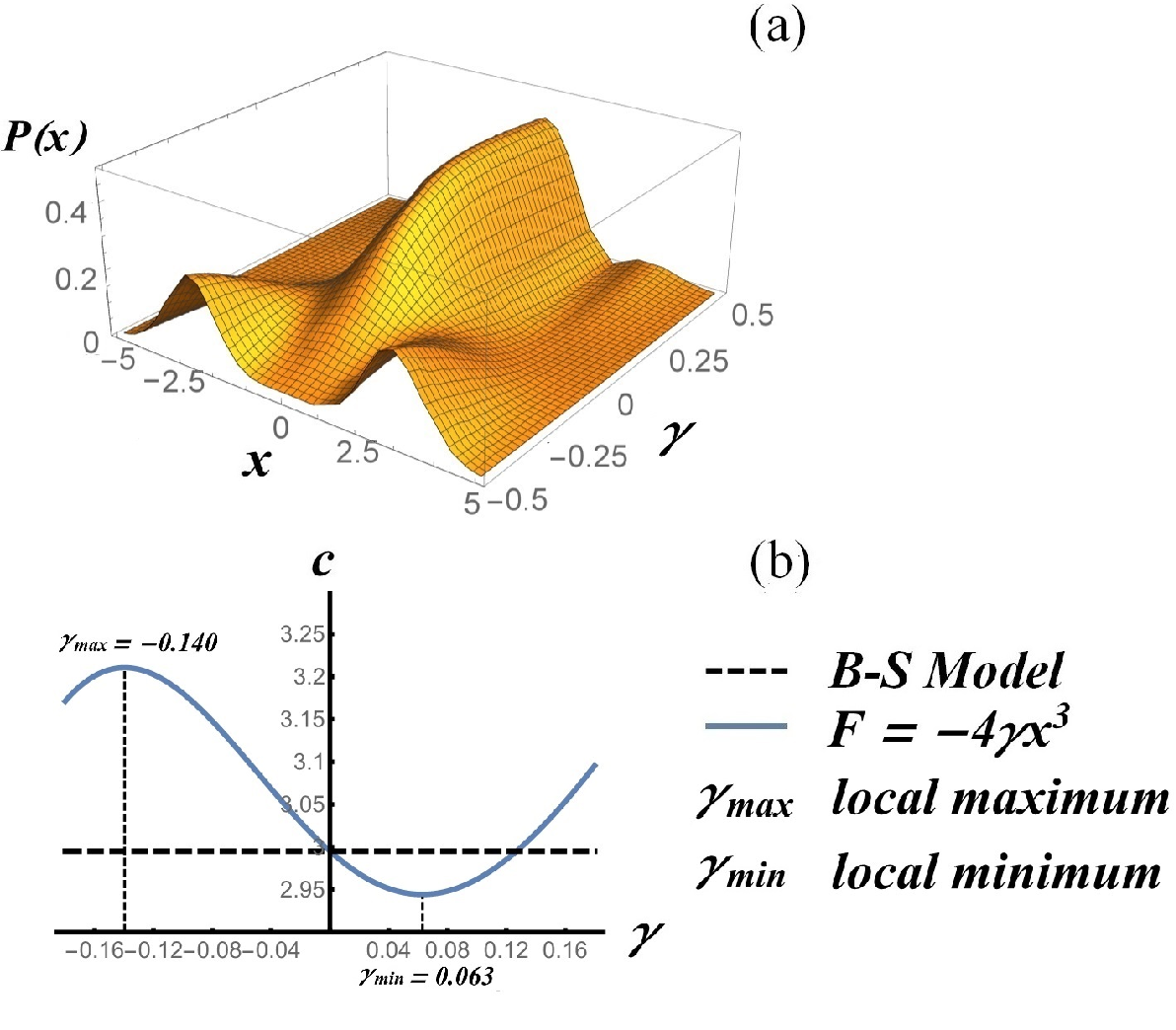}
	\caption{(Color online)(a)
		The probability density versus $\gamma$
		When $\gamma=0$ it reduces the normal distribution assumed in Black-Scholes Model.
		(b) The call option price versus $\gamma$.
		The plot we use $r=10\%$, $S_0=20$, $K=20$, $T=1year$ and $\sigma=25\%$.
	}
	\label{fig4}
\end{figure}

Similarly for small $\gamma$, by the perturbation method we can obtain the approximate
solution of wave function in the ground state. The solution of wave function in the first-order approximation can be given by (see Appendix)
\begin{equation}\label{WFP4}
	\psi_{g}(x)=\frac{C}{(2\pi)^{1/4}}e^{-x^2/4}\left[1-\frac{\gamma}{4\sqrt{2}\hbar\omega}(x^4-9)\right]
\end{equation}
The probability density is obtained
\begin{equation}\label{PDP4}
	P_{\beta}(x)=C^2\frac{e^{-x^2/2}}{\sqrt{2\pi}}
	\left[1-\frac{\gamma}{4\sqrt{2}\hbar\omega}(x^4-9)\right]^{2}
\end{equation}
where $C$ is the normalization constant.
We plot the probability density versus $\gamma$ in Fig. $\ref{fig4}$(a) and the option price in Fig. $\ref{fig4}$(b).

It should be noticed that the finance market depends on both $\gamma$ and $x$ shown in Fig.$\ref{fig4}$(a) .
For $\gamma >0$, $F<0$ means that the force is resistant for $x>0$ and the force is active  for  $x<0$. For $\gamma <0$, $F>0$ means that the force is active for $x<0$ and the force is resistent for $x>0$. As $|\lambda|$ increases, the original price could be unstable and there exist two symmetric attractors, which push the price either up or down.
Fig. $\ref{fig4}$(b) shows the call option price versus $\gamma$. The call option price shows a minimum at $\gamma\approx 0.063$ and a maximum at $\gamma\approx -0.14$.
When the price distribution becomes less homogeneous, the call option price will be lower. This perfectly matches the realistic situation that the less fluctuate price of underlying assets leads to a lower call option price.

Furthermore, we can also apply this method to solve any other polynomial boundary conditions. For an arbitrary condition, we can expand the function with Taylor series and we can analyze the change of the call option price to an arbitrary hidden market forces.

\subsection{Quantum well}

In finance market, if a company does not want its Underlying Asset price lower than $S_n$, it is equivalent to exist a boundary condition that makes the dealing of its asset stop.
From quantum mechanics point of views we may set up a quantum well model
to simulate this behavior.
The potential of quantum well is
\begin{equation}\label{QW1}
	V(x)=\left\{
	\begin{array}{c}
		0 \quad for \quad |x|< a \\
		\infty \quad for \quad |x|\geq a
	\end{array}
	\right.
\end{equation}
where $a\propto \Delta S$. It implies when $x\leq X_n$ ($X_n\propto S_n-S_0$)
the finance market is in the normal state and the market will stop if some unexpected forces make the Underlying Asset price going beyond the lower or upper bound, namely lower than $S_0-\Delta S$ or  higher than $S_0+\Delta S$. $|a|$ can be regarded as a boundary of finance market.

The solution of wave function in ground state is obtained
\begin{equation}\label{QWwf1}
	\psi_{g}(x)=\frac{1}{\sqrt{a}}\sin\left[{\frac{\pi}{2a}}(x+a)\right],\quad x\in[-a,+a]
\end{equation}
and the probability density is expressed as
\begin{equation}\label{QWP}
	P_{QW}(x)=\left\{
	\begin{array}{l}
		\frac{1}{a}\sin^2\left[{\frac{\pi}{2a}}(x+a)\right],\quad x\in[-a,+a]  \nonumber\\
		0, \qquad \qquad \qquad \qquad x\in(-\infty,-a)\bigcup(a,+\infty)
	\end{array}
	\right.
\end{equation}

\begin{figure}[t]
	\centering
	\includegraphics[scale=0.45]{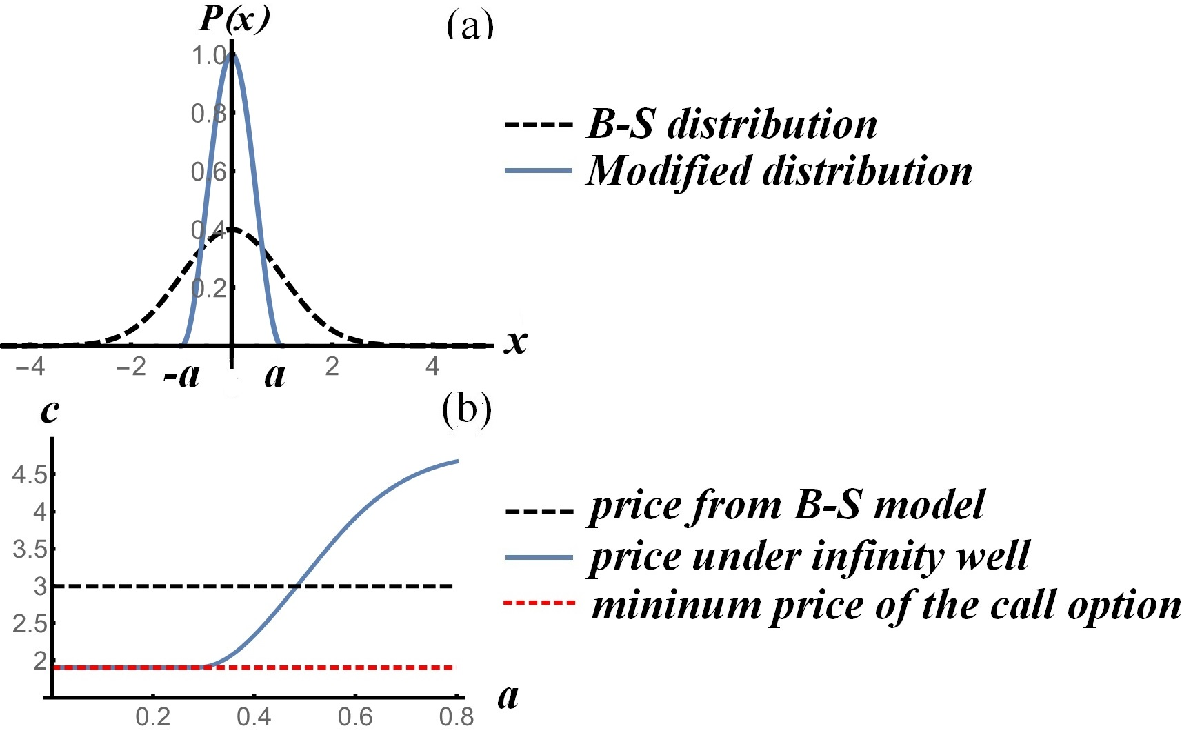}
	\caption{(Color online)(a) Comparison of Probability Densities between the normal distribution and distribution in quantum well. $\bm{P(x)}$:
		The half width of modified distribution is the half width between two barriers (infinity wells). Practical meaning of a is the width of price range for the underlying asset.
		(b) The call option price change along with $a$.
		We use $r=10\%$, $S_0=20$, $K=20$, $T=1year$ and $\sigma=25\%$.
		If there is no arbitrary chance, call option price should be no less than $S_0-Ke^ {-rT}$ which is regarded as the minimum price for call options.}
	\label{fig5}
\end{figure}

Since the boundary of the well prevents neither the underlying price increasing too very high nor decreasing too very low, the whole distribution will be squeezed within $-a<x<a$. The probability of the future underlying price still near current price significantly increases, but when $x\rightarrow\pm a$, the probability decreased below the normal distribution seen in Fig. $\ref{fig5}$(a).
By plugging the distribution function in Eq. ({\ref{QWP}}), we can plot the $c$ versus $\lambda$ curve for the call option price with $a$. The figure $\ref{fig5}$(b) shows that when $a<0.3$, the original price does not change. The call option price is equal to $S_0-Ke^ {-eT}$. As $a$ increases the call option price will also increase. This result shows that as the price distribution becomes more homogeneous, the call option will be more expensive, which matches the realistic situation.

As illustrative examples here we present the call option pricing based on this framework. Similarly we can also give the modification of the put option pricing in the practical application.

\section{Conclusion}
The Black-Scholes model gives a scheme of option pricing based on the risk-free,
efficient market hypothesis and the standard Guassian stochastic dynamics of finance market.
In the realistic finance market there exist various unpredictable factors, such
as abnormal shorting or buying some assets, some rules or policy changes, some
unexpected news and some psychological features of investors, which could break
the efficient market hypothesis making the stochastic dynamics deviate
from the standard Guassian stochastic dynamics of finance market. How do we
take these factors into account to generalize the Black-Scholes model for the option pricing becomes a practical problem. We discover the analog between the Guassian stochastic dynamics
and the probability density of the ground state of quantum harmonic oscillator. We propose a market force model to simulate various unpredictable market behaviors to modifying
the Guassian dynamics based on quantum mechanics approach. Based on this model
we give a new scheme of option pricing for various unpredictable market behaviors.
The option pricing based on quantum mechanics provides a more systematic and explicit
way to generalize the Guassian probability distribution for the Black-Scholes model
than other approaches.\cite{09}

As examples, we present several market forces to generalize the Black-Schole model
and turn out their corresponding option pricing. In principle, we can generalize further this method to more complicated forces driving the finance market because any form of forces
or potentials as a function of $x$ can be expanded by Taylor series. By the perturbation method we can calculate arbitrary-order approximation of wave function based on the quantum mechanics approach. On the other hand, we can also generalize the one-dimensional oscillator to the multi-dimensional oscillators, which covers various market forces
and their interactions which could modify the option pricing.

This study on the option pricing provides two practical hints. One is that as a new scheme
of option pricing we can modify the pure Black-Scholes model-based option pricing when we can predict some unexpected force emergence. The other is that when
one cannot predict these unexpected forces appearing. The prediction of the option pricing
based on this scheme provides some risk premium.

%\section*{Declarations}

%Some journals require declarations to be submitted in a standardised format. Please check the Instructions for Authors of the journal to which you are submitting to see if you need to complete this section. If yes, your manuscript must contain the following sections under the heading `Declarations':

\section{Appendices}

\subsection{The quantum harmonic oscillator}

The Hamiltonian of 1D harmonic oscillator is written as

\begin{equation}
	H=-\frac{\hbar^{2}}{2m}\frac{d^{2}}{dx^{2}}+\frac{1}{2}m\omega^{2}x^{2}
\end{equation}
The wave function can be obtained

\begin{equation}
	\psi _{n}(x) =\left( \frac{\alpha }{\sqrt{\pi }2^{n}n!}\right)
	^{1/2}e^{-\alpha ^{2}x^{2}/2}H_{n}(x)
\end{equation}
where $\alpha =\sqrt{\frac{m\omega }{\hbar }}\equiv \frac{1}{\sqrt{2}}$ for convenience.
The eigen energy is

\begin{equation}
	E_{n} =\hbar \omega \left( n+\frac{1}{2}\right)
\end{equation}
where $n =0,1,2,\cdots$.

\subsection{The perturbation method}

The energy and wave function by the non-degenerate perturbation are

\begin{eqnarray}
	E_{n} &=&E_{n}^{(0)}+H_{nn}^{\prime }+\sum'_{m}\frac{|H_{nm}^{\prime }|^{2}}{%
		E_{n}^{(0)}-E_{m}^{(0)}}+... \\
	\psi _{n} &=&\psi _{n}^{(0)}+\sum'_{m}\frac{H_{nm}^{\prime }}{%
		E_{n}^{(0)}-E_{m}^{(0)}}\psi _{m}^{(0)\prime }+...  \label{Awf}
\end{eqnarray}
where%
\begin{equation}
	H_{nm}^{\prime }=\langle \psi _{n}|H^{\prime }|\psi _{m}\rangle
\end{equation}
is the matrix element of the perturbation Hamiltonian.

\subsection{The perturbation wave function in the ground state of harmonic oscillator}
The ground state wave function of harmonic oscillator is
\begin{equation}
	\psi _{0}(x) =\left( \frac{\alpha }{\sqrt{\pi }}\right)
	^{1/2}e^{-\alpha ^{2}x^{2}/2}
\end{equation}
For convenience, we introduce the occupation number representation
\begin{eqnarray}
	\widehat{a} &=&\frac{\alpha }{\sqrt{2}}\left( x+\frac{i}{\alpha ^{2}}\widehat{p}_{x}\right) \\
	\widehat{a}^{\dagger } &=&\frac{\alpha }{\sqrt{2}}\left( x-\frac{i}{\alpha ^{2}}%
	\widehat{p}_{x}\right)
\end{eqnarray}
with
\begin{eqnarray}
	x &=&\frac{1}{\sqrt{2}\alpha }\left( \widehat{a}+\widehat{a}^{\dagger }\right) \\
	\widehat{p}_{x} &=&\frac{\alpha }{i\sqrt{2}}\left( \widehat{a}-\widehat{a}^{\dagger }\right)
\end{eqnarray}
The commutative relation is $\left[\widehat{a},\widehat{a}^{\dagger } \right]=1$ and
\begin{eqnarray}
	\widehat{a}|n\rangle &=&\sqrt{n}|n-1\rangle \\
	\widehat{a}^{\dagger }|n\rangle &=&\sqrt{n+1}|n+1\rangle
\end{eqnarray}

\textbf{Case I}: $x^{2}$ forces

The perturbation Hamiltonian is
\begin{equation}
	H^{\prime }=\beta x^{3}=\frac{\beta }{2\sqrt{2}\alpha ^{3}}\left(
	\widehat{a}+\widehat{a}^{\dagger }\right) ^{3}
\end{equation}
and the perturbation matrix element in the ground state is expressed as
\begin{equation}
	H_{0m}^{\prime } =\langle 0|H^{\prime }|m\rangle
	=\frac{\beta }{2\sqrt{2}\alpha ^{3}}\langle 0|\left( a+a^{\dagger }\right)
	^{3}|m\rangle
\end{equation}
By using formula in Eq. ($\ref{Awf}$), we can obtain
\begin{equation}
	\psi _{0} =C\left( \frac{1}{\sqrt{2\pi }}\right) ^{1/2}e^{-x^{2}/4}
	\left[ 1-\frac{\beta }{\hbar \omega }\left( \frac{1}{3}x^{3}+\frac{x}{2}\right) \right]
\end{equation}
The probability density can be expressed as
\begin{equation}
	P_{\beta }(x)=\frac{C^{2}e^{-x^{2}/2}}{\sqrt{2\pi }}\left[ 1-\frac{\beta }{\hbar \omega }\left( \frac{1}{3}x^{3}+\frac{x}{2}\right) \right] ^{2}
\end{equation}

\textbf{Case II}: $x^{3}$ forces

The perturbation Hamiltonian is
\begin{equation}
	H^{\prime }=\gamma x^{4}=\frac{\gamma }{4\alpha ^{4}}\left( a+a^{\dagger
	}\right) ^{4}
\end{equation}
and the perturbation matrix element in the ground state is expressed as
\begin{equation}
	H_{0m}^{\prime } =\langle 0|H^{\prime }|m\rangle
	=\frac{\gamma }{4\alpha ^{4}}\langle 0|\left( a+a^{\dagger }\right)
	^{4}|m\rangle
\end{equation}
Similarly, we can obtain
\begin{equation}
	\psi _{0} =C\left( \frac{1}{\sqrt{2\pi }}\right) ^{1/2}e^{-x^{2}/4}\left[ 1-\frac{%
		\gamma}{4\sqrt{2}\hbar \omega}\left( x^{4}-9\right) \right]
\end{equation}
The probability density is
\begin{equation}
	P_{\gamma }(x)=\frac{C^{2}e^{-x^{2}/2}}{\sqrt{2\pi }}
	\left[ 1-\frac{\gamma }{4\sqrt{2}\hbar \omega}
	\left(x^{4}-9\right) \right] ^{2}
\end{equation}

	% =========================
	% 参考文献
	% =========================
%	\bibliographystyle{unsrt}
%	\bibliography{references}
\bibliography{apssamp}

\begin{thebibliography}{99}
\bibitem{01} {F.Black and M.Scholes},{The Pricing of Options and Corporate Liabilities},{Journal of Political Economy},{\bf 637-59},(May/June, 1973).

\bibitem{02} {J. C.Hull},{Options,futures, and other Derivatives the 10th Edition},
{\bf 320-323}, Pearson, (2011).

\bibitem{03} {B. E. Baaquie},{Quantum Finance}, {\bf 316}, 78-115 (2004)
\bibitem{04} {Lauterbach, Beni and Schultz, Paul}, {Journal of Finance},{\bf 45}, 1181-1209 (1990).
\bibitem{05} {Scott, Louis O.}, {Option Pricing when the Variance Changes Randomly: Theory, Estimation, and an Application}, {\bf 22}, 419-438 (1987).
\bibitem{06} {Kleinert, H. and Korbel, J.}, {Physica A Statistical Mechanics \& Its Applications}, {\bf 449}, 200-214 (2016).
\bibitem{07} {Song, Lina and Wang, Weiguo}, {Abstract \& Applied Analysis}, {\bf 2013}, 1-16(2013).
\bibitem{08} {Merton, Robert C.},{J Financial Economics}, {\bf 3}, 125-144 (1976).
\bibitem{09} {Mantegna, Rosario Nunzio},{Physica A Statistical Mechanics \& Its Applications} ,{\bf 179}, 232 (1991).
\bibitem{10} {B. E. Baaquie},{Quantum Finance},{\bf 316}, 76-77 (2004).
\bibitem{11} {B. E. Baaquie},{Quantum Finance}, {\bf 316}, 73-76 (2004).
	
\end{thebibliography}
% Produces the bibliography via BibTeX.

\end{document}